%% file: acmart.tex
\documentclass[sigconf]{acmart}

\AtBeginDocument{%
  }

\copyrightyear{2024}
\acmYear{2024}
\setcopyright{acmlicensed}\acmConference[ICSE-SEIP '24]{46th International
Conference on Software Engineering: Software Engineering in Practice}{April
14--20, 2024}{Lisbon, Portugal}
\acmBooktitle{46th International Conference on Software Engineering: Software
Engineering in Practice (ICSE-SEIP '24), April 14--20, 2024, Lisbon, Portugal}
\acmDOI{10.1145/3639477.3639727}
\acmISBN{979-8-4007-0501-4/24/04}

\input{packages}
\input{macros}

\allowdisplaybreaks

\begin{document}

\title{Translating between SQL Dialects for Cloud Migration}

\author{Ran Zmigrod}
\email{ran.zmigrod@jpmchase.com}
\orcid{0009-0001-0169-1670}
\affiliation{%
  \institution{J.P. Morgan AI Research}
  \city{London}
  \country{UK}
}
\author{Salwa Alamir}
\email{salwa.alamir@jpmchase.com}
\orcid{0009-0006-6650-7041}
\affiliation{%
  \institution{J.P. Morgan AI Research}
  \city{London}
  \country{UK}
}
\author{Xiaomo Liu}
\email{xiaomo.liu@jpmchase.com}
\orcid{0000-0003-4184-4202}
\affiliation{%
  \institution{J.P. Morgan AI Research}
  \state{New York}
  \country{USA}
}

\renewcommand{\shortauthors}{Zmigrod et al.}

\begin{abstract}
  Migrations of systems from on-site premises to the cloud has been a fundamental endeavor by many industrial institutions.
  A crucial component of such cloud migrations is the transition of databases to be hosted online.
  In this work, we consider the difficulties of this migration for SQL databases.
  While SQL is one of the prominent methods for storing database procedures, there are a plethora of different SQL dialects (e.g., MySQL, Postgres, etc.) which can complicate migrations when the on-premise SQL dialect differs to the dialect hosted on the cloud.
  Tools exist by common cloud provides such as AWS and Azure to aid in translating between dialects in order to mitigate the majority of the difficulties.
  However, these tools do not successfully translate $100\%$ of the code.
  Consequently, software engineers must manually convert the remainder of the untranslated database.
  For large organizations, this task quickly becomes intractable and so more innovative solutions are required.
  We consider this challenge a novel yet vital industrial research problem for any large corporation that is considering cloud migrations.
  Furthermore, we introduce potential avenues of research to tackle this challenge that have yielded promising preliminary results.
\end{abstract}



\maketitle

\input{figure}

\section{Introduction}
The interest and demand of migration to the cloud has been accelerating over the past decade.
Indeed, more and more companies are choosing to move away from on-premise capabilities in favor of hosting applications and resources on the cloud \cite{jamshidi-13-cloud, shuaib-19-why, gleb-21-systematic}.
An integral part to these efforts is SQL database migration to the cloud \cite{iqbal-19-key, mukherjee-19-benefits}.
This can often be a simple task when on-site and cloud infrastructures align such as when the on-premise SQL dialect matches the cloud SQL dialect.
However, when this is not the case, companies may need to heavily invest in tools, frameworks, and developer hours to address this problem.

Large cloud service providers 
often do not support all SQL-based dialects and so translations between SQL dialects are necessary.
Schema conversion tools 
\cite{aws-sct, azure-dsct}
exist to largely automate this task.
Unfortunately, these tools do not always convert $100\%$ of code and so developers are left to manually convert any segments of code that the tools could not.
While this may not be overwhelming for smaller migrations, larger migrations can involve hundreds of thousands of lines of SQL code and would require significant work even if only $1\%$ of the code could not be converted.
Therefore, industrial practitioners must seek more automated methods \cite{alamir-22-updates, navarro-23-update} to efficiently migrate databases to the cloud.\looseness=-1

In this paper, we introduce a crucial industrial challenge in database migration to the research community: How can we automatically transform SQL segments that state-of-the-art database conversion tools fail on?
Due to the lack of available public and private data for this problem, many modern solutions that capitalize on machine learning approaches \cite{roziere-20-unsupervised, jiang-21-cure, dehaerne-22-code} are not applicable.
We thus propose three avenues that may be worth exploring for tackling the challenge: Manual rule creation, imitation learning (IL), and large language models (LLMs).

\section{Problem Definition}
The challenge this work introduces is more complex than SQL-to-SQL translation and is best explained alongside \cref{fig:example}.
Consider a migration of a database hosted with {\color{ColorSQL1Text} SQL dialect 1} into a database hosted with {\color{ColorSQL2Text} SQL dialect 2}.
As a first step, we utilize common tools 
\cite{azure-dsct, aws-sct}
to convert the majority of the database.
However, these tools are often unable to convert the entire code base and unconverted segments must be dealt with manually; unconverted segments are typically accompanied by a {\color{ColorErrorText} conversion error} and tend to be more complex than code segments that were successfully converted.
The task is thus to automatically transform the {\color{ColorSQL1Text} SQL dialect 1} segment and its corresponding {\color{ColorErrorText} error} into an {\color{ColorSQL2Text} SQL dialect 2} segment to complete the migration.
This problem is critical for industry as large manual efforts cause serious bottlenecks for cloud migrations.
These bear large financial impacts such as dual hosting applications and databases on-premise and on the cloud as well as the developer costs of performing the manual conversions.

Access to data is one of the largest hurdles to overcome in this challenge.
Large organizations typically cannot publish data for open research due to confidentiality issues and thus research must rely on public data for this task.
Unfortunately, no public dataset currently exists for this problem.
While it may be possible to curate a dataset through web scraping,
it is unlikely this endeavor will correctly capture the task at hand.
This is because we are specifically tackling transformations that could not be handled by common tools and have an error indicating the issue.
Therefore, we would not know the correct fix to the extracted dataset.
Additionally, publicly available code often does not represent the complexities of industry code \cite{heitlager-07-practical} and will likely be convertable using existing tools.
Therefore, we must consider solutions to this challenge using low data (or even no data) environments rather than common translation solutions that utilize machine learning \cite{roziere-20-unsupervised, jiang-21-cure, dehaerne-22-code}.

\section{Possible Approaches}
In this section we propose three methods that practitioners and researchers may use to tackle the SQL transformation challenge: Manual rule creation, IL, and leveraging LLMs.\footnote{We further note that if the scope is limited to very simple cases, we can reduce the SQL migration problem to a programming language abstraction problem.
}

\subsection{Manual Rule Creation}
The majority of SQL parsers are rule-based.
This is due to the finite syntax and deterministic grammar of SQL dialects.
While details of
widely used conversion tools
are not publicly available, through analysis of their outputs (specifically conversion errors), we believe that these popular conversion tools are also rule-based.
Therefore, perhaps the simplest approach for targeting our problem is to devise rules for re-occurring error examples.
Such rules can then be integrated on top of converter tools to perform a complete database migration.
This method has one clear limitation: We must manually construct rules.
Not only is this time consuming, but it requires both knowledge of the SQL dialects in question as well as the conversion mechanics (i.e., how are rules implemented).
As such, while applying rules has been the predominant method for SQL parsers and converters, we do not believe it is a sustainable solution for the task at hand.

\subsection{Imitation Learning}
Imitation learning \cite{hussein-17-imitation} is a branch of reinforcement learning, an artificial intelligence technique used to train systems to follow a given ``policy'', and has been adopted in a plethora of different contexts \cite{berant-15-imitation, hussein-17-imitation, codevilla-18-end}.
In our case, the policy entails how to convert an SQL segment from one dialect to another given an error.
IL specifically requires an ``expert'' (in our case, a developer familiar with the necessary manual conversion) to provide the learning system with an example of what should be done, and the system consequently learns to imitate this behavior.

Similar to rule creation, this method requires that we have knowledge of how to preform the manual conversions that the initial tools could not.
However, IL does not require us to develop the rules ourselves.
Furthermore, only a handful of examples (sometimes even just a single example) of each error is needed for the system to be able to create a rule, meaning that minimal work is required for highly re-occurring conversion errors.
Consequently, both the barrier of skill as well as the effort required to use an IL tool by a cloud migration team is much lower than the manual rule creation necessitates.
Indeed, we developed an IL tool that treats SQL segments as trees and so conversion rules are tree transforms.
The tool successfully learned to handle over $80\%$ of an initial test set\footnote{We collated $11$ simple and unique conversion errors. Our IL tool was able to correctly convert nine of these after being supplied one or two examples of each error. Due to space constraints, we do not provide further details of our system.} and is being further developed and tested with more complex errors.

\subsection{Large Language Models}
LLMs, such as GPT-4 \cite{gpt4}, Bard \cite{bard}, \emph{inter alia}, have demonstrated impressive code writing capabilities \cite{ouyang-23-llm}, thereby presenting themselves as interesting and promising solutions for this challenge.
Perhaps the most appealing quality of using LLMs for these translations, is that no manual conversions must be done.
However, there is no guarantee that a solution produced by an LLM is correct.
Research has demonstrated that LLMs are prone to hallucinations \cite{weisz-21-perfection, liu-23-your, ouyang-23-llm}, meaning that there may be syntactic or semantic errors in any SQL segments generated.
Through initial experimentation with GPT-3.5, we did find that along with correct conversions, certain errors (specifically those of greater complexity) yielded incorrect conversions.
Therefore, any solution that applies LLMs must include robust verification that the returned segment is grammatical with respect to the target SQL dialect and that it preserves the semantics of the original SQL segment.

\section{Conclusion}
In this paper, we introduced a unique problem to large SQL
migrations that are common throughout industry.
Companies must currently plan unsustainable large manual conversion efforts of SQL segments that widely-used tools cannot currently handle.
Therefore, organizations are faced with a difficult and urgent research challenge to automate the process.
We outlined three avenues to tackle this challenge: Manual rule creation, IL, and LLMs.
Each solution enables improved automation and both IL and LLMs are showing promising results.
We hope that this paper inspires further work on the solutions presented and perhaps encourages altogether new solutions to this challenge.

\section*{Disclaimer}
This paper was prepared for informational purposes by the Artificial Intelligence Research group of JPMorgan Chase \& Co and its affiliates (“JP Morgan”), and is not a product of the Research Department of JP Morgan. JP Morgan makes no representation and warranty whatsoever and disclaims all liability, for the completeness, accuracy or reliability of the information contained herein. This document is not intended as investment research or investment advice, or a recommendation, offer or solicitation for the purchase or sale of any security, financial instrument, financial product or service, or to be used in any way for evaluating the merits of participating in any transaction, and shall not constitute a solicitation under any jurisdiction or to any person, if such solicitation under such jurisdiction or to such person would be unlawful.

\bibliographystyle{ACM-Reference-Format}
\bibliography{references}

\end{document}

%% file: packages.tex
\usepackage{enumitem}
\usepackage{relsize}
\usepackage{booktabs}
\usepackage{tabularx}
\usepackage{multirow}
\usepackage{graphicx}
\usepackage{xspace}
\usepackage{xcolor}
\usepackage{caption}
\usepackage{subcaption}
\usepackage{listings}
\usepackage{soul}
\usepackage{cleveref}

\usepackage{amsmath}
\usepackage{amsthm}
\usepackage{amsfonts}
\usepackage{mathtools}

\usepackage{tikz}
\usetikzlibrary{arrows.meta}
\usetikzlibrary{shapes.arrows}
\usetikzlibrary{shapes.misc}
\usetikzlibrary{fit}
\usetikzlibrary{calc,shapes}
\usetikzlibrary{tikzmark}
\usetikzlibrary{arrows}
\usetikzlibrary{trees}

%% file: macros.tex
\crefname{chapter}{Chapter}{}
\crefname{section}{Section}{Sections}
\crefformat{section}{Section~#2#1#3}

\crefname{table}{Table}{Tables}
\crefname{figure}{Figure}{Figures}
\crefname{algorithm}{Alg.}{Algs.}
\crefname{line}{Line}{Lines}
\crefname{appendix}{App.}{}
\crefname{chapter}{Chapter}{Chapters}

\crefname{thm}{Theorem}{Theorems}
\crefname{prop}{Proposition}{Propositions}
\crefname{definition}{Definition}{Definitions}
\crefname{lemma}{Lemma}{Lemmas}
\crefname{cor}{Corollary}{Corollaries}
\crefname{equation}{Eq.}{Eqs.}

\creflabelformat{equation}{#2\textup{#1}#3}
\crefrangelabelformat{equation}{(#3#1#4--#5#2#6)}

\theoremstyle{definition}

\newcolumntype{C}{>{\centering\arraybackslash}X}

\renewcommand{\ldots}{\ensuremath{{\ldotp\kern-0.2em\ldotp\kern-0.2em\ldotp}}}

\renewcommand{\cdots}{\ensuremath{{\cdotp\kern-0.2em\cdotp\kern-0.2em\cdotp}}}
\renewcommand{\dots}{\ensuremath{{\ldotp\kern-0.2em\ldotp\kern-0.2em\ldotp}}}

\definecolor{ColorSQL1}{rgb}{0.96,0.72,0.38}
\definecolor{ColorSQL2}{rgb}{0.64,0.86,0.54}
\definecolor{ColorError}{rgb}{0.93,0.6,0.52}
\definecolor{ColorConverter}{rgb}{0.49,0.63,0.87}
\definecolor{ColorSQL1Text}{rgb}{0.79,0.54,0.19}
\definecolor{ColorSQL2Text}{rgb}{0.36,0.62,0.32}
\definecolor{ColorErrorText}{rgb}{0.8,0.15,0}
\definecolor{ColorConverterText}{rgb}{0.31,0.45,0.73}

%% file: figure.tex
\lstset{escapeinside={(*@}{@*)}}
\begin{figure}[t]
    \centering
    \begin{lstlisting}[
           language=SQL,
           showspaces=false,
           basicstyle=\ttfamily,
           numbers=left,
           numberstyle=\tiny,
           commentstyle=\color{gray},
           numbers=none,
           frame=single,
           morekeywords={DECLARE},
           backgroundcolor = \color{ColorSQL1!30},
           rulecolor=\color{ColorSQL1Text}
        ]
    DECLARE var1 VARCHAR(20) = NULL
    SELECT (*@\hl{var1}@*) + "string"
    AS var2
    \end{lstlisting}
    \begin{tikzpicture}
        \node (recth) at (-4,0) [draw,thick,minimum width=0.6cm,minimum height=0.05cm, fill=black] {};
        \node (rectv) at (-4,0) [draw,thick,minimum height=0.6cm,minimum width=0.05cm, fill=black] {};
        \node (error) at (0.5, 0) [fill=ColorError!30, minimum width=7.4cm, minimum height=1.cm, draw=ColorErrorText] {};
        \node at (0.5, 0) {\begin{tabular}{l}  Error: String concatenation between NULL  and NOT \\ NULL values makes the whole string AS NULL\end{tabular}};
        \node (rectangle) at (0, -1.1) [fill=ColorConverter!30, minimum width=2.4cm, minimum height=0.7cm] {};
        \fill[fill=ColorConverter!30] (-1.5,-1.4) node[anchor=north]{}
          -- (1.5,-1.4) node[anchor=north]{}
          -- (0,-1.8) node[anchor=south]{}
          -- cycle;
        \node[text width=3cm,align=center] at (0, -1.1) {{\bf SQL Converter}};
    \end{tikzpicture}
    \begin{lstlisting}[
           language=SQL,
           showspaces=false,
           basicstyle=\ttfamily,
           numbers=left,
           numberstyle=\tiny,
           commentstyle=\color{gray},
           numbers=none,
           frame=single,
           morekeywords={DECLARE},
           backgroundcolor = \color{ColorSQL2!30},
           rulecolor=\color{ColorSQL2Text}
        ]
    DECLARE var1 VARCHAR(20) DEFAULT NULL
    SELECT CONCAT((*@\hl{\textbf{ISNULL}(var1, "")}@*), "string")
    AS var2
    \end{lstlisting}
    
    \caption{Example of SQL transformation challenge. The aim is for the {\color{ColorConverterText} SQL converter} to transform an {\color{ColorSQL1Text} SQL dialect 1 segment} with a {\color{ColorErrorText} conversion error} into an {\color{ColorSQL2Text} SQL dialect 2 segment}.}
    \label{fig:example}
\end{figure}

%% file: acmart.bbl

\begin{thebibliography}{21}


\ifx \showCODEN    \undefined \def \showCODEN     #1{\unskip}     \fi
\ifx \showDOI      \undefined \def \showDOI       #1{#1}\fi
\ifx \showISBNx    \undefined \def \showISBNx     #1{\unskip}     \fi
\ifx \showISBNxiii \undefined \def \showISBNxiii  #1{\unskip}     \fi
\ifx \showISSN     \undefined \def \showISSN      #1{\unskip}     \fi
\ifx \showLCCN     \undefined \def \showLCCN      #1{\unskip}     \fi
\ifx \shownote     \undefined \def \shownote      #1{#1}          \fi
\ifx \showarticletitle \undefined \def \showarticletitle #1{#1}   \fi
\ifx \showURL      \undefined \def \showURL       {\relax}        \fi
\providecommand\bibfield[2]{#2}
\providecommand\bibinfo[2]{#2}
\providecommand\natexlab[1]{#1}
\providecommand\showeprint[2][]{arXiv:#2}

\bibitem[Alamir et~al\mbox{.}(2022)]%
        {alamir-22-updates}
\bibfield{author}{\bibinfo{person}{Salwa Alamir}, \bibinfo{person}{Petr Babkin}, \bibinfo{person}{Nacho Navarro}, {and} \bibinfo{person}{Sameena Shah}.} \bibinfo{year}{2022}\natexlab{}.
\newblock \showarticletitle{{AI} for Automated Code Updates}. In \bibinfo{booktitle}{\emph{44th {IEEE/ACM} International Conference on Software Engineering: Software Engineering in Practice, {ICSE} {(SEIP)} 2022, Pittsburgh, PA, USA, May 22-24, 2022}}. \bibinfo{publisher}{{IEEE}}, \bibinfo{pages}{25--26}.
\newblock
\urldef\tempurl%
\url{https://doi.org/10.1109/ICSE-SEIP55303.2022.9794071}
\showDOI{\tempurl}


\bibitem[Amazon({[n.\,d.]})]%
        {aws-sct}
\bibfield{author}{\bibinfo{person}{Amazon}.} \bibinfo{year}{[n.\,d.]}\natexlab{}.
\newblock \bibinfo{title}{Amazon Web Services Schema Conversion Tool}.
\newblock
\newblock
\urldef\tempurl%
\url{https://aws.amazon.com/dms/schema-conversion-tool/}
\showURL{%
\tempurl}


\bibitem[Berant and Liang(2015)]%
        {berant-15-imitation}
\bibfield{author}{\bibinfo{person}{Jonathan Berant} {and} \bibinfo{person}{Percy Liang}.} \bibinfo{year}{2015}\natexlab{}.
\newblock \showarticletitle{Imitation Learning of Agenda-based Semantic Parsers}.
\newblock \bibinfo{journal}{\emph{Trans. Assoc. Comput. Linguistics}}  \bibinfo{volume}{3} (\bibinfo{year}{2015}), \bibinfo{pages}{545--558}.
\newblock
\urldef\tempurl%
\url{https://doi.org/10.1162/tacl\_a\_00157}
\showDOI{\tempurl}


\bibitem[Codevilla et~al\mbox{.}(2018)]%
        {codevilla-18-end}
\bibfield{author}{\bibinfo{person}{Felipe Codevilla}, \bibinfo{person}{Matthias M{\"{u}}ller}, \bibinfo{person}{Antonio~M. L{\'{o}}pez}, \bibinfo{person}{Vladlen Koltun}, {and} \bibinfo{person}{Alexey Dosovitskiy}.} \bibinfo{year}{2018}\natexlab{}.
\newblock \showarticletitle{End-to-End Driving Via Conditional Imitation Learning}. In \bibinfo{booktitle}{\emph{2018 {IEEE} International Conference on Robotics and Automation, {ICRA} 2018, Brisbane, Australia, May 21-25, 2018}}. \bibinfo{publisher}{{IEEE}}, \bibinfo{pages}{1--9}.
\newblock
\urldef\tempurl%
\url{https://doi.org/10.1109/ICRA.2018.8460487}
\showDOI{\tempurl}


\bibitem[Dehaerne et~al\mbox{.}(2022)]%
        {dehaerne-22-code}
\bibfield{author}{\bibinfo{person}{Enrique Dehaerne}, \bibinfo{person}{Bappaditya Dey}, \bibinfo{person}{Sandip Halder}, \bibinfo{person}{Stefan~De Gendt}, {and} \bibinfo{person}{Wannes Meert}.} \bibinfo{year}{2022}\natexlab{}.
\newblock \showarticletitle{Code Generation Using Machine Learning: {A} Systematic Review}.
\newblock \bibinfo{journal}{\emph{{IEEE} Access}}  \bibinfo{volume}{10} (\bibinfo{year}{2022}), \bibinfo{pages}{82434--82455}.
\newblock
\urldef\tempurl%
\url{https://doi.org/10.1109/ACCESS.2022.3196347}
\showDOI{\tempurl}


\bibitem[Gleb(2021)]%
        {gleb-21-systematic}
\bibfield{author}{\bibinfo{person}{Taras Gleb}.} \bibinfo{year}{2021}\natexlab{}.
\newblock \bibinfo{booktitle}{\emph{Cloud Migration Fundamentals}}.
\newblock \bibinfo{publisher}{Apress}, \bibinfo{address}{Berkeley, CA}, \bibinfo{pages}{19--35}.
\newblock
\showISBNx{978-1-4842-7252-7}
\urldef\tempurl%
\url{https://doi.org/10.1007/978-1-4842-7252-7_2}
\showDOI{\tempurl}


\bibitem[Heitlager et~al\mbox{.}(2007)]%
        {heitlager-07-practical}
\bibfield{author}{\bibinfo{person}{Ilja Heitlager}, \bibinfo{person}{Tobias Kuipers}, {and} \bibinfo{person}{Joost Visser}.} \bibinfo{year}{2007}\natexlab{}.
\newblock \showarticletitle{A Practical Model for Measuring Maintainability}. In \bibinfo{booktitle}{\emph{Quality of Information and Communications Technology, 6th International Conference on the Quality of Information and Communications Technology, {QUATIC} 2007, Lisbon, Portugal, September 12-14, 2007, Proceedings}}, \bibfield{editor}{\bibinfo{person}{Ricardo~Jorge Machado}, \bibinfo{person}{Fernando~Brito e~Abreu}, {and} \bibinfo{person}{Paulo~Rupino da~Cunha}} (Eds.). \bibinfo{publisher}{{IEEE} Computer Society}, \bibinfo{pages}{30--39}.
\newblock
\urldef\tempurl%
\url{https://doi.org/10.1109/QUATIC.2007.8}
\showDOI{\tempurl}


\bibitem[Hussein et~al\mbox{.}(2017)]%
        {hussein-17-imitation}
\bibfield{author}{\bibinfo{person}{Ahmed Hussein}, \bibinfo{person}{Mohamed~Medhat Gaber}, \bibinfo{person}{Eyad Elyan}, {and} \bibinfo{person}{Chrisina Jayne}.} \bibinfo{year}{2017}\natexlab{}.
\newblock \showarticletitle{Imitation Learning: {A} Survey of Learning Methods}.
\newblock \bibinfo{journal}{\emph{{ACM} Comput. Surv.}} \bibinfo{volume}{50}, \bibinfo{number}{2} (\bibinfo{year}{2017}), \bibinfo{pages}{21:1--21:35}.
\newblock
\urldef\tempurl%
\url{https://doi.org/10.1145/3054912}
\showDOI{\tempurl}


\bibitem[Iqbal and Palacios(2019)]%
        {iqbal-19-key}
\bibfield{author}{\bibinfo{person}{Arif Iqbal} {and} \bibinfo{person}{Ricardo~Colomo Palacios}.} \bibinfo{year}{2019}\natexlab{}.
\newblock \showarticletitle{Key Opportunities and Challenges of Data Migration in Cloud: Results from a Multivocal Literature Review}. In \bibinfo{booktitle}{\emph{{CENTERIS} 2019 - International Conference on ENTERprise Information Systems / ProjMAN 2019 - International Conference on Project MANagement / HCist 2019 - International Conference on Health and Social Care Information Systems and Technologies 2019, Sousse, Tunisia}} \emph{(\bibinfo{series}{Procedia Computer Science}, Vol.~\bibinfo{volume}{164})}, \bibfield{editor}{\bibinfo{person}{Maria~Manuela Cruz{-}Cunha}, \bibinfo{person}{Ricardo Martinho}, \bibinfo{person}{Rui Rijo}, \bibinfo{person}{Emanuel Peres}, {and} \bibinfo{person}{Dulce Domingos}} (Eds.). \bibinfo{publisher}{Elsevier}, \bibinfo{pages}{48--55}.
\newblock
\urldef\tempurl%
\url{https://doi.org/10.1016/j.procs.2019.12.153}
\showDOI{\tempurl}


\bibitem[Jamshidi et~al\mbox{.}(2013)]%
        {jamshidi-13-cloud}
\bibfield{author}{\bibinfo{person}{Pooyan Jamshidi}, \bibinfo{person}{Aakash Ahmad}, {and} \bibinfo{person}{Claus Pahl}.} \bibinfo{year}{2013}\natexlab{}.
\newblock \showarticletitle{Cloud Migration Research: {A} Systematic Review}.
\newblock \bibinfo{journal}{\emph{{IEEE} Trans. Cloud Comput.}} \bibinfo{volume}{1}, \bibinfo{number}{2} (\bibinfo{year}{2013}), \bibinfo{pages}{142--157}.
\newblock
\urldef\tempurl%
\url{https://doi.org/10.1109/TCC.2013.10}
\showDOI{\tempurl}


\bibitem[Jiang et~al\mbox{.}(2021)]%
        {jiang-21-cure}
\bibfield{author}{\bibinfo{person}{Nan Jiang}, \bibinfo{person}{Thibaud Lutellier}, {and} \bibinfo{person}{Lin Tan}.} \bibinfo{year}{2021}\natexlab{}.
\newblock \showarticletitle{{CURE:} Code-Aware Neural Machine Translation for Automatic Program Repair}. In \bibinfo{booktitle}{\emph{43rd {IEEE/ACM} International Conference on Software Engineering, {ICSE} 2021, Madrid, Spain, 22-30 May 2021}}. \bibinfo{publisher}{{IEEE}}, \bibinfo{pages}{1161--1173}.
\newblock
\urldef\tempurl%
\url{https://doi.org/10.1109/ICSE43902.2021.00107}
\showDOI{\tempurl}


\bibitem[Liu et~al\mbox{.}(2023)]%
        {liu-23-your}
\bibfield{author}{\bibinfo{person}{Jiawei Liu}, \bibinfo{person}{Chunqiu~Steven Xia}, \bibinfo{person}{Yuyao Wang}, {and} \bibinfo{person}{Lingming Zhang}.} \bibinfo{year}{2023}\natexlab{}.
\newblock \showarticletitle{Is Your Code Generated by ChatGPT Really Correct? Rigorous Evaluation of Large Language Models for Code Generation}.
\newblock \bibinfo{journal}{\emph{CoRR}}  \bibinfo{volume}{abs/2305.01210} (\bibinfo{year}{2023}).
\newblock
\urldef\tempurl%
\url{https://doi.org/10.48550/arXiv.2305.01210}
\showDOI{\tempurl}
\showeprint[arXiv]{2305.01210}


\bibitem[Manyika(2023)]%
        {bard}
\bibfield{author}{\bibinfo{person}{James Manyika}.} \bibinfo{year}{2023}\natexlab{}.
\newblock \showarticletitle{An overview of Bard: an early experiment with generative {AI}}.
\newblock \bibinfo{journal}{\emph{Google}} (\bibinfo{year}{2023}).
\newblock
\urldef\tempurl%
\url{https://ai.google/static/documents/google-about-bard.pdf}
\showURL{%
\tempurl}


\bibitem[Microsoft({[n.\,d.]})]%
        {azure-dsct}
\bibfield{author}{\bibinfo{person}{Microsoft}.} \bibinfo{year}{[n.\,d.]}\natexlab{}.
\newblock \bibinfo{title}{Microsoft Azure Database Schema Conversion Toolkit}.
\newblock
\newblock
\urldef\tempurl%
\url{https://learn.microsoft.com/en-us/sql/azure-data-studio/extensions/dsct/database-schema-conversion-toolkit}
\showURL{%
\tempurl}


\bibitem[Mukherjee(2019)]%
        {mukherjee-19-benefits}
\bibfield{author}{\bibinfo{person}{Sourav Mukherjee}.} \bibinfo{year}{2019}\natexlab{}.
\newblock \showarticletitle{Benefits of {AWS} in Modern Cloud}.
\newblock \bibinfo{journal}{\emph{CoRR}}  \bibinfo{volume}{abs/1903.03219} (\bibinfo{year}{2019}).
\newblock
\showeprint[arXiv]{1903.03219}
\urldef\tempurl%
\url{http://arxiv.org/abs/1903.03219}
\showURL{%
\tempurl}


\bibitem[Navarro et~al\mbox{.}(2023)]%
        {navarro-23-update}
\bibfield{author}{\bibinfo{person}{Nacho Navarro}, \bibinfo{person}{Salwa Alamir}, \bibinfo{person}{Petr Babkin}, {and} \bibinfo{person}{Sameena Shah}.} \bibinfo{year}{2023}\natexlab{}.
\newblock \showarticletitle{An Automated Code Update Tool For Python Packages}. In \bibinfo{booktitle}{\emph{2023 IEEE International Conference on Software Maintenance and Evolution (ICSME)}}. \bibinfo{pages}{536--540}.
\newblock
\urldef\tempurl%
\url{https://doi.org/10.1109/ICSME58846.2023.00068}
\showDOI{\tempurl}


\bibitem[OpenAI(2023)]%
        {gpt4}
\bibfield{author}{\bibinfo{person}{OpenAI}.} \bibinfo{year}{2023}\natexlab{}.
\newblock \showarticletitle{{GPT-4} Technical Report}.
\newblock \bibinfo{journal}{\emph{CoRR}}  \bibinfo{volume}{abs/2303.08774} (\bibinfo{year}{2023}).
\newblock
\urldef\tempurl%
\url{https://doi.org/10.48550/arXiv.2303.08774}
\showDOI{\tempurl}
\showeprint[arXiv]{2303.08774}


\bibitem[Ouyang et~al\mbox{.}(2023)]%
        {ouyang-23-llm}
\bibfield{author}{\bibinfo{person}{Shuyin Ouyang}, \bibinfo{person}{Jie~M. Zhang}, \bibinfo{person}{Mark Harman}, {and} \bibinfo{person}{Meng Wang}.} \bibinfo{year}{2023}\natexlab{}.
\newblock \showarticletitle{{LLM} is Like a Box of Chocolates: {T}he Non-determinism of ChatGPT in Code Generation}.
\newblock \bibinfo{journal}{\emph{CoRR}}  \bibinfo{volume}{abs/2308.02828} (\bibinfo{year}{2023}).
\newblock
\urldef\tempurl%
\url{https://doi.org/10.48550/arXiv.2308.02828}
\showDOI{\tempurl}
\showeprint[arXiv]{2308.02828}


\bibitem[Rozi{\`{e}}re et~al\mbox{.}(2020)]%
        {roziere-20-unsupervised}
\bibfield{author}{\bibinfo{person}{Baptiste Rozi{\`{e}}re}, \bibinfo{person}{Marie{-}Anne Lachaux}, \bibinfo{person}{Lowik Chanussot}, {and} \bibinfo{person}{Guillaume Lample}.} \bibinfo{year}{2020}\natexlab{}.
\newblock \showarticletitle{Unsupervised Translation of Programming Languages}. In \bibinfo{booktitle}{\emph{Advances in Neural Information Processing Systems 33: Annual Conference on Neural Information Processing Systems 2020, NeurIPS 2020, December 6-12, 2020, virtual}}, \bibfield{editor}{\bibinfo{person}{Hugo Larochelle}, \bibinfo{person}{Marc'Aurelio Ranzato}, \bibinfo{person}{Raia Hadsell}, \bibinfo{person}{Maria{-}Florina Balcan}, {and} \bibinfo{person}{Hsuan{-}Tien Lin}} (Eds.).
\newblock
\urldef\tempurl%
\url{https://proceedings.neurips.cc/paper/2020/hash/ed23fbf18c2cd35f8c7f8de44f85c08d-Abstract.html}
\showURL{%
\tempurl}


\bibitem[Shuaib et~al\mbox{.}(2019)]%
        {shuaib-19-why}
\bibfield{author}{\bibinfo{person}{Mohammed Shuaib}, \bibinfo{person}{Abdus Samad}, \bibinfo{person}{Shadab Alam}, {and} \bibinfo{person}{Shams~T. Siddiqui}.} \bibinfo{year}{2019}\natexlab{}.
\newblock \showarticletitle{Why Adopting Cloud Is Still a Challenge?---A Review on Issues and Challenges for Cloud Migration in Organizations}. In \bibinfo{booktitle}{\emph{Ambient Communications and Computer Systems}}, \bibfield{editor}{\bibinfo{person}{Yu-Chen Hu}, \bibinfo{person}{Shailesh Tiwari}, \bibinfo{person}{Krishn~K. Mishra}, {and} \bibinfo{person}{Munesh~C. Trivedi}} (Eds.). \bibinfo{publisher}{Springer Singapore}, \bibinfo{address}{Singapore}, \bibinfo{pages}{387--399}.
\newblock
\showISBNx{978-981-13-5934-7}


\bibitem[Weisz et~al\mbox{.}(2021)]%
        {weisz-21-perfection}
\bibfield{author}{\bibinfo{person}{Justin~D. Weisz}, \bibinfo{person}{Michael~J. Muller}, \bibinfo{person}{Stephanie Houde}, \bibinfo{person}{John~T. Richards}, \bibinfo{person}{Steven~I. Ross}, \bibinfo{person}{Fernando Martinez}, \bibinfo{person}{Mayank Agarwal}, {and} \bibinfo{person}{Kartik Talamadupula}.} \bibinfo{year}{2021}\natexlab{}.
\newblock \showarticletitle{Perfection Not Required? Human-AI Partnerships in Code Translation}. In \bibinfo{booktitle}{\emph{{IUI} '21: 26th International Conference on Intelligent User Interfaces, College Station, TX, USA, April 13-17, 2021}}, \bibfield{editor}{\bibinfo{person}{Tracy Hammond}, \bibinfo{person}{Katrien Verbert}, \bibinfo{person}{Dennis Parra}, \bibinfo{person}{Bart~P. Knijnenburg}, \bibinfo{person}{John O'Donovan}, {and} \bibinfo{person}{Paul Teale}} (Eds.). \bibinfo{publisher}{{ACM}}, \bibinfo{pages}{402--412}.
\newblock
\urldef\tempurl%
\url{https://doi.org/10.1145/3397481.3450656}
\showDOI{\tempurl}


\end{thebibliography}
